\documentclass[epj]{svjour}
\usepackage{graphics}
\newcommand{\be}{\begin{equation}}
\newcommand{\ee}{\end{equation}}
\newcommand{\ba}{\begin{eqnarray}}
\newcommand{\ea}{\end{eqnarray}}
\newcommand{\bb}{\begin{thebibliography}}
\newcommand{\eb}{\end{thebibliography}}
\newcommand{\ci}[1]{\cite{#1}}
\newcommand{\bi}[1]{\bibitem{#1}}
\newcommand{\lab}[1]{\label{#1}}

\begin{document}

%
\title{Is there a hadron spin-flip contribution \\
  to the Coulomb-hadron interference
  at small transfer momenta and high energies
   }
\titlerunning{Is there exist a hadron spin-flip contribution  ...}
\author{ O.V. Selyugin 
}                     
%
%
\institute{
 BLTPh, JINR, Dubna, Russia
}
\date{Received: date / Revised version: date}
%

\abstract{
   The analysing power $A_N$  is examined in the range of the
  Coulomb-hadron interference on the basis of the
 experimental data from $p_L = 6 \ GeV/c$ up to $200 \ GeV/c$
    taking account  of a phenomenological analysis at $p_L = 6 \
    GeV/c$ and  a dynamic high-energy spin model.
The results are
    compared with the new RHIC data at  $p_L = 100 \ GeV/c$.
     The new experimental data obtained at RHIC indicate
     small contributions of the hadron spin-flip amplitude.
} 
\maketitle
\section{Introduction}
\label{intro}

  Most of the recent experiments require the knowledge of the polarization
    of beams with high accuracy.
  This especially  relates to the  large spin programs at
  RHIC.  These programs include  measurements of the spin
 correlation parameters in the diffraction range of elastic
 proton-proton scattering.
 There is a proposal to use the Coulomb-
 nucleon interference (CNI) effects \ci{swin}
 to measure very exactly and fast
 the beam polarization \ci{sum}.
 This effect appears from the interference of the
 imaginary part of the hadron non-spin-flip amplitude and the real part
 of the electromagnetic spin-flip amplitude determined by the
  charge-magnetic moment interaction.
  Now the new very precise experimental data are obtained at RHIC
  \cite{Bravar1,Bravar2}.

     Determination of the structure of the hadron scattering amplitude
 is an important task for both the theory and experiment.
 Perturbative Quantum Chromodynamics cannot be used
 in  calculation of the real and imaginary
 parts of the scattering amplitude in the diffraction range.
 A worse situation is for the spin-flip parts of the scattering
 amplitude in the domain of small transfer momenta. On the one hand,
 the usual representation says  that the spin-flip amplitude  dies
 at superhigh energies, and, on the other hand, we have  different
 non-perturbative models which lead to a non-dying spin-flip amplitude
 at superhigh energies \ci{bsw,mog1,zpc}.

    Note that the interference of the hadronic
 and electromagnetic amplitudes may give an important contribution not
 only at very small transfer momenta but also in the range of the
 diffraction minimum \cite{selpht}.  However, for that one should know the
 phase of the interference of the Coulombic and hadronic amplitude at
 sufficiently large transfer momenta too.

  Before the RHIC experiments experimental data on the measurement
   of the spin correlation parameters at very small transfer momenta were
  very poor except
  the unique experiment \ci{akex} though with large errors.
  After the first paper \ci{akth}  a number of papers  appeared
  which considered these questions and tried to estimate a possible
  contribution of the hadron-spin-flip amplitude to the CNI effect
  \ci{trosh,soff,btt96}.

Our difficulty is  mostly defined by the
 lack of  experimental data at high energies and small transfer
 momenta.  We should examine the available experimental data at
 different energies and in different domains of transfer momenta. In
  most analyses  the experimental data at $p_L = 45.5 \ GeV/c$
  and with $0.06 < |t| < 0.5 \ GeV^2$ and the data at $p_L=200 \ GeV/c$
  with $0.003 < |t| < 0.05$ are used. These experimental data overlap
  on the axis of transfer momenta but are measured at different
  energies.
         In most analyses the energy difference of all parameters determining
  the scattering amplitude  is
  not considered.  Of course, we have plenty of  experimental
  data in the  domain of small transfer momenta at low energies $3
  < p_L < 12 \ (GeV/c)$.


  At these energies we have many contributions
   to the hadron-spin-flip amplitudes coming from
  different regions of exchange.
  In \cite{berg} note that at not-high energies the Reggions $\rho$ and $A_2$ give
  a dominant contributions in the hadron spin-flip amplitude.
   However,  Regge-pole-exchange contributions occur in exchange-degenerate
   pairs, their imaginary parts cancelling.
  Now we cannot exactly calculate all
  contributions and find their energy dependence. However, a great amount of
  the experimental material allows us to make full
  phenomenological analyses, and obtain the size and form of the
  different parts of the hadron scattering amplitude.  The difficulty
  is that we do not know the energy dependence of these amplitudes and
  individual contributions of the asymptotic non-dying spin-flip
  amplitudes. As was noted in \cite{wak}, the spin-dependent part
  of the interaction in $pp$ scattering is stronger than was expected and a good
  fit to the data in the Regge model requires an enormous number of poles.

  Usually, one takes the
  assumptions that the imaginary and real parts of the spin-non-flip
  amplitude have the exponential behavior with the same slope and the
  imaginary and real parts of the spin-flip amplitudes, without the
  kinematic factor $\sqrt{|t|}$.
 For example, in \ci{sum-L}
  the spin-flip amplitude was   chosen in  the form
 \ba
       F_{h}^{fl}=\sqrt{-t}/(2m_{p}) (R_5 + i I_5) Im F_{h}^{nf}.
\ea
  That is not so simple as regards the $t$ dependence
  shown in Ref. \ci{soff}, where
  $F^{fl}_{h}$ multiply the exponential form by the special function
   dependent on $t$.
  Moreover, one mostly  takes  the energy independence of
  the ratio of the spin-flip parts to the spin-non-flip parts of the
  scattering amplitude.  All this is our theoretical uncertainty \cite{M-Pred}.

\section{Model approximation}

 In \cite{cnizdr}, the phenomenological analysis of the experimental
 data was carried out to estimate
 the size of the hadron spin-flip amplitude from the experimental
 data on differential cross sections,  the
 influence of the hadron-spin flip amplitude on the CNI effect
 and a possibility
 of  estimating this contribution from the experimental data
 on measurement
 of the analyzing power in the nucleon-nucleon elastic scattering.
   Now we can compare those results with the new experimental data
   obtained at RHIC.

   The differential cross sections measured in an experiment
 are described by the square of the scattering amplitude
   which is used to fit   experimental  data
  determining the electromagnetic
  and hadron amplitudes and the Coulomb-hadron phase.

   For the electromagnetic helicity amplitudes, one  takes the usual
     one-photon approximations (see \ci{gaz,pred})
\begin{eqnarray}
 F^{em}_{1,3}(t) &=& \frac{\alpha}{t} \ f_1(t)^2; \nonumber \\
 F^{em}_2(t)&=&-F^{em}_4(t) = \alpha \ f_2^2(t); \nonumber \\
 F^{em}_5(t) &=& - \frac{\alpha}{ \sqrt{|t|}} \ f_1(t) \  f_2(t).
\end{eqnarray}
 with
\ba
f_1(t)&=& \frac{ 4 \ m_p^2 - (\mu_p - 1) \ t}{4 \ m_p^2 - t} \ G_D;  \nonumber \\
f_2(t)&=&\frac{2 \ m_p \ (\mu_p - 1) }{4 \ m_p^2 - t} \ G_D; \nonumber \\
G_D(t)& = &  \frac{1}{1-t/0.71^2}; \ \ \  \nonumber \\
  \mu_p &=&  2.793; \ \ \ m_p = 0.93827 \ GeV.
\ea

 As a result,  the total helicity amplitudes can be written
 as
 \ba
   F_i(s,t) = F_i^H(s,t) + F_i^{em}(t) e^{-i \alpha  \varphi(s,t)},
\ea
 with
 the Coulomb-hadron phase \ci{selpht} calculated for the whole diffraction
 range with taking into account the hadron form-factors.
 The differential cross
  sections and spin correlation parameters are
 \ba
  \frac{d\sigma}{dt}&=& 2 \pi(|F_1|^2+|F_2|^2+|F_3|^2+|F_4|^2+4|F_5|^2).
                                \lab{dsdt}
\ea
\ba
  A_N\frac{d\sigma}{dt}&=& -4\pi Im[(F_1+F_2+F_3-F_4)*F_5^{*}).  \lab{an}
\ea
   We shell restrict our discussion to the analysis of  $A_N$.
   In the standard pictures the spin-flip and double spin-flip amplitudes
    correspond to the spin-orbit $(LS)$ and spin-spin $(SS)$ coupling terms.
  The contribution to
  $A_N$ from the hadron double spin-flip amplitudes
   already at $p_L = 6 \ $GeV/c is of the second order
  compared to the contribution from spin-flip amplitude.
   So, with the usual high energy approximation for the helicity amplitudes
   at  small transfer momenta we suppose that
   $F_{1}=F_{3}$ and we can neglect the contributions of the hadron parts
   of $F_2-F_4$.
   Note that if $F_1, F_3, F_5$ have the same phases, their interference contribution
   to $A_N$ will be zero, though the size of the hadron spin-flip amplitude can be large.
   Hence, if this phases has a different $s$ and $t$ dependence, the contribution from the hadron
   spin-flip amplitude in $A_N$ can be zero at $s_i, \ t_i$ and non-zero at other $s_j, \ t_j$.
   It means that the comparison of the size $A_N(s)$ at one $t_i$, as made for example
    in \cite{tru-an-t}, at different $s$ has the strong
   assumption about energy independence many different parameters determining the size of $A_N(s,t)$.

   The analysing power corresponding
   to the pure electro-magnetic-hadron interference
   (with $F_5^H =0$)  will be
   denoted by $A_N^{CH}$.
   Its size is proportional, in major part, to the interference of the imaginary part of the
   hadron spin-non-flip amplitude with the real part of the electromagnetic spin-flip
    amplitude.
   Note that there is also a small contribution from
   the interference of the real and imaginary part of the above mentioned amplitudes.

    The existing experimental data at sufficiently high energy shows the significant size of
    $A_N$ in the $t$-region of the dip of the differential cross sections.
   At the present moment, we have, as has been noted above,  that in
  some models the hadron asymptotical spin-flip amplitude is not dying
  at super-high energy.  However, most part of the experimental data of the
  analyzing power lies at low energies. Hence, we should take the low
  energy amplitudes and build a continues transition to the
  asymptotic amplitudes.

    As asymptotic amplitudes let us take those calculated
  in the dynamical model (DM)  \cite{zpc}.
  In  \cite{g2}  on
  the basis of sum rules it has been shown that the main contribution  to
  a hadron interaction at large  distances  comes  from  the  triangle
  diagram with the $2\pi $ -meson exchange in the $t$-channel. As a
  result, the hadron amplitude can be  represented  as  a  sum  of
  central and peripheral parts of the interaction
\ba F(s,t) \propto
F_{c}(s,t) + F_{p}(s,t),
\ea
where $F_{c}(s,t)$   describes   the
interaction between the central parts of hadrons; and   $F_{p}(s,t)$
is the sum of contributions of diagrams corresponding to the
interactions of the central part of one hadron  on  the  meson cloud of
the other. The contribution of these diagrams to the scattering
amplitude  with  an $N(\Delta $-isobar)  in  the intermediate state
looks like \ci{zpc}
\ba
 && F ^{\lambda _{1}\lambda _{2}}_{N(\Delta )}(s,t)=
{g^{2}_{\pi NN(\Delta )}\over i(2\pi )^{4}}\int
 d^{4}q F_{\pi N}(s\acute{,}t)  \nonumber \\
\times && \frac{ \varphi _{N(\Delta )}
[(k-q),q^{2}]\varphi _{N(\Delta )}[(p-q),q^{2}]}
{[q^{2}- M^{2}_{N(\Delta )}+ i\epsilon ]}           \nonumber \\
\times && \frac{\Gamma^{\lambda_{1}\lambda_{2}}(q,p,k,)}{[(k-q)^{2}- \mu ^{2}+i\epsilon ]
[(p-q)^{2}- \mu^{2}+ i\epsilon ]}.
\ea
Here $\lambda_{1}$ and $\lambda_{2}$  are the helicities  of  nucleons;
$F_{\pi N}$  is  the
$\pi N$-scattering amplitude; $\Gamma$  is a matrix element of the numerator
of the diagram representation; $\varphi $ are vertex functions chosen
 in the dipole form  with
the parameters $\beta _{N(\Delta )}$:
\ba
\varphi _{N(\Delta )}(l^{2},q^{2}\propto  M^{2}_{N(\Delta )})
= {\beta ^{4}_{N(\Delta )}\over (\beta ^{2}_{N(\Delta )}- l^{2})^{2}}.
\ea

     The model with  the $N $  and  $\Delta $
contribution  provides  a self-consistent  picture  of
the differential cross sections and spin phenomena
of different hadron processes  at  high  energies.
Really, parameters in the  amplitude  determined from,
 for example, elastic $pp$-scattering, allow one to  obtain
a wide range of results for elastic
meson-nucleon scattering and charge-exchange reaction
 $\pi^{-} p \rightarrow  \pi^{0} n$
 at high energies.

 It is essential that  the
model predicts  large  polarization  effects  for  all  considered
reactions at high and superhigh energies \ci{zpc}. The predictions  are  in
good agreement with the experimental data  in  the  energy  region
available for experiment. Also note that just the effect of  large
distances determines a large value of the  spin-flip  amplitude  of
the charge-exchange reaction \ci{g4}.

 The results  weakly depend  on the model for the spin-non-flip
amplitude. Different models must give the same differential cross
 sections in a wide range of transfer momenta and energies.
 Moreover, they must describe the energy dependence of $\rho(s) = Re
 F(s,0)/Im F(s,0)$.  Basically, only the behavior of the real part of
the spin-non-flip amplitudes in the range of the  diffraction minimum may
 depend on the model and leads to different predictions.
 In this paper, we consider a usual picture of the proton-proton
 and proton-antiproton cross
 sections with the crossing symmetry fulfilled.

    As a low energy amplitude let us take the one obtained
   in  \ci{wak}
  where the full analysis of  experimental data was
 carried out and the full set of the helicity spin amplitudes
 and their eikonals of the proton-proton scattering at $p_L = 6 \ $GeV/c
  was extracted.
  Let us take the eikonal of  the spin-non-flip  amplitudes in the form
 similar to the form and size                       obtained in
\ci{wak} at $p_L = 6 \ GeV/c $  :
    \ba 1- e^{\chi_{c}(b)} & = & h_{1} e^{-c_{1} b^{2}} - h_{2}
  e^{-c_{2} b^{2}} + h_{3} e^{-c_{3} b^{2}}   \nonumber \\ & & +i \
   (h_{4} e^{-c_{4} b^{2}} - h_{5} e^{-c{5} b^{2}} + h_{6} e^{- c_{6}
       b^{2}})                           \lab{wak1}
  \ea
  and for the hadron spin-flip amplitude
  \ba \chi_{ls}(b) = h_{ls} [1+ b \ e^{\mu(s)
    (b - b_{0}) }]^{-1},    \lab{wak2}
    \ea
  where $h_{i}$,$c_{i}$, $h_{ls}$ and $b_{0}$
  are the parameters obtained in Ref. \ci{wak}.
  As we know, these amplitudes  reproduce the analyzing power
 at $p_L = 6 \ GeV/c$.
  In fact, these amplitudes are a sum of terms falling, constant and
  growing with energy.
   However, this form has no energy dependence of the parameters
 which change the form of these amplitudes with increasing energy in
 both the spin-non-flip and spin-flip parts. To obtain the energy
  dependence of some part of the amplitude (\ref{wak1}, \ref{wak2}),
  let us multiply (\ref{wak2}) by the falling term $s_1/s$ and  take
  into account the change of the form of (\ref{wak2}) with energy;  let
  us introduce the energy dependence into the parameter $\mu
  \rightarrow \mu_{s}$
\ba
      \mu(s) = \mu_{0} ( \log{s_{0}} / \log{s} ),
\ea
 where $s_0 = 13.152 \ $GeV  corresponds to $p_L = 6 \ $GeV/c and $\mu_{0}$
  corresponds to the values of Ref. \ci{wak}.

  The DM amplitude also includes the falling, constant, and
  increasing terms, but it is not suitable for describing
  low-energy data. So this is not a simple task to sew these two
  amplitudes, low energy phenomenological and high energy model
 amplitudes.  To obtain a smooth transform to the DM
 representation, let us multiply these amplitudes by
   the factor-functions  $ fs^{nf,fl}_{ex} $ quickly
 decreasing with energy, and multiply
 the DM amplitudes by the factor-functions $ f^{nf,fl}_{th} $
  \ba
 fs^{nf}_{ex}(s) &=& exp[-(s/s^{nf})^{2}+(s_0/s^{nf})^2]; \nonumber \\
 \ \ fs^{nf}_{th}(s) &=& 1 - exp[-(s/s^{nf})^{2}+(s_0/s^{nf})^2];
 \ea
 \ba fs^{fl}_{ex}(s) &=&
  exp[-(s/s^{fl})^{2}+(s_0/s^{fl})^2]; \nonumber \\
 \ \ fs^{nf}_{th}(s) &=& 1 - exp[-(s/s^{fl})^{2}+(s_0/s^{fl})^2],
\ea
where $s_0 = 13.152 \ $GeV is correspond to $p_L = 6 \ $GeV/c.
  In this case, we obtain that the analyzing power at $p_L = 6 \ GeV/c$
  is described only by the amplitudes obtained in Ref. \ci{wak} and at
  superhigh energies only by the DM amplitude. In the domain of
 approximately $6 \leq p_L \leq 200 ( GeV/c)$  the analyzing power has
 both the contributions.  The parameters $s^{nf}$ and $s^{fl}$ were chosen
  to obtain the description of experimental data available in this
  energy range: $  s^{nf} = 40 \ GeV^2; \ \ \  s^{fl} = 64 \ GeV^2  $.
   We do not carry out the fitting procedure. The values of these parameters
   were chosen to obtain a qualitative description of the polarization data
   at $p_L = 11.75 \ $GeV/c.

   We do not take into account the data of
   the differential cross sections. However, to check up our procedure
   we calculate the differential cross sections at $p_L =50 \ $GeV/c
   and  at $p_L =100 \ $GeV/c and compare them with the existing
   experimental data.
\begin{figure*}
\resizebox{0.85\textwidth}{!}{%
 \phantom{.} \hspace{30mm} \includegraphics{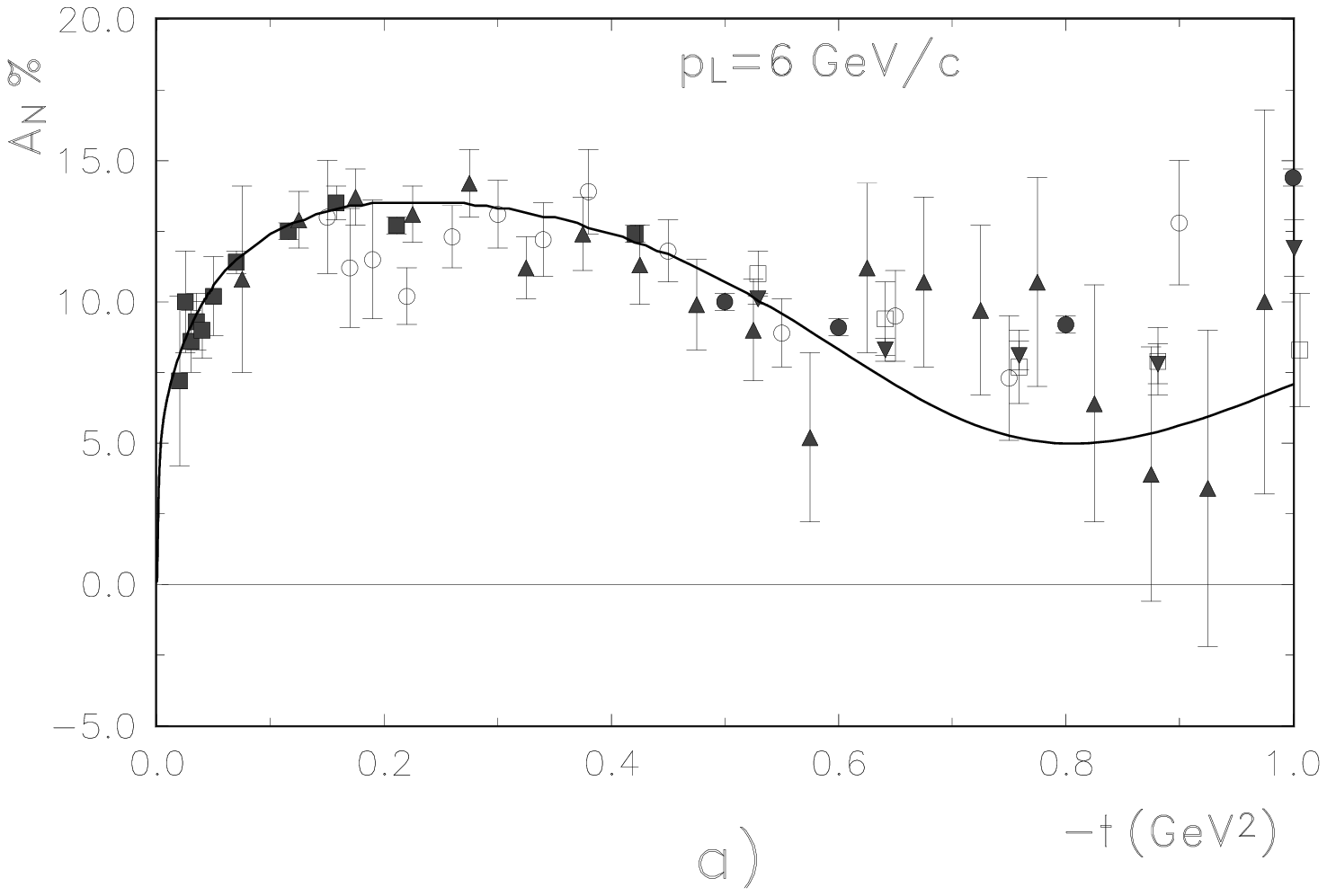}
  \includegraphics{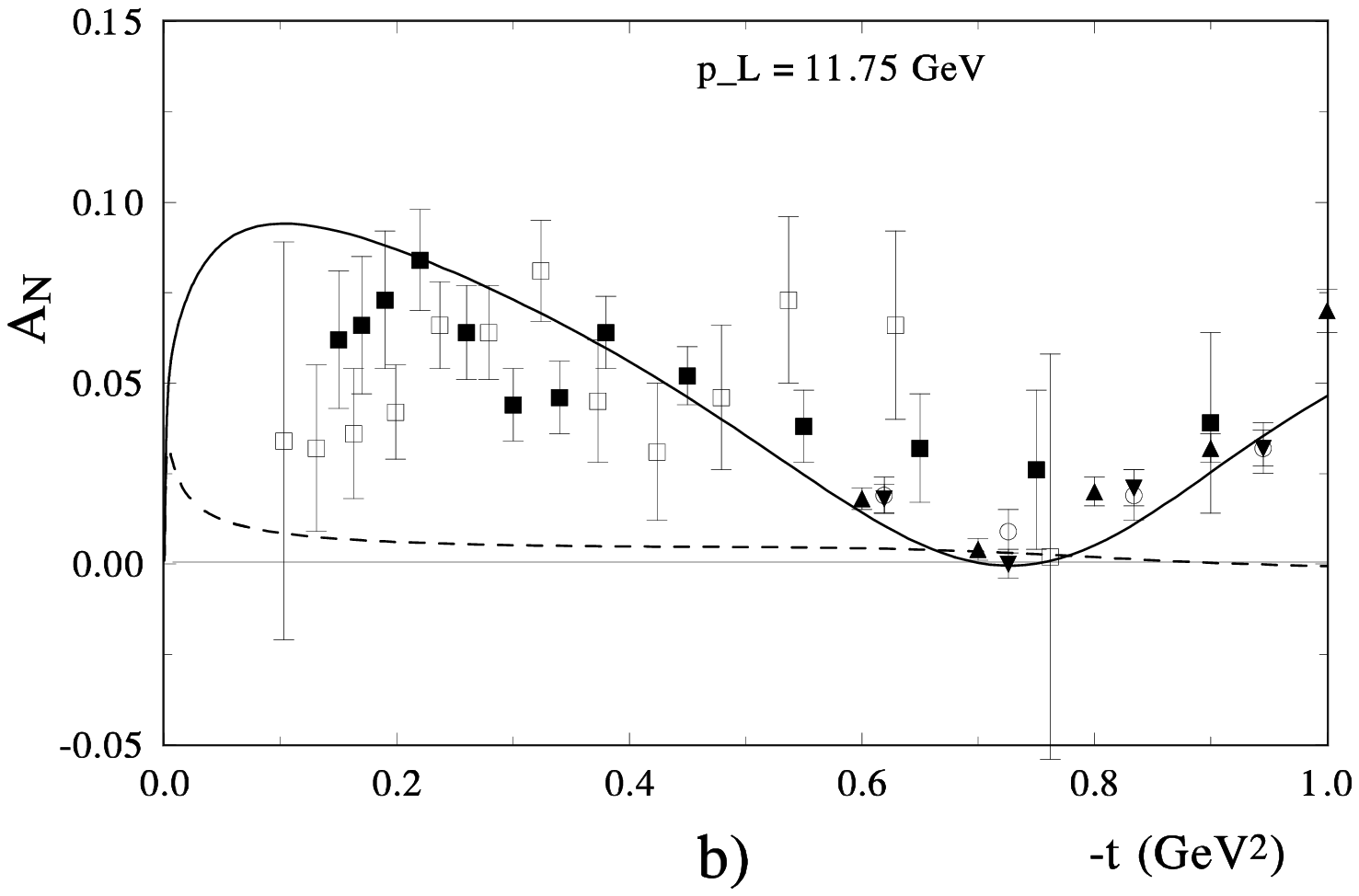}
}
\caption{The analyzing power $A_N$ of pp - scattering
      calculated:
  a) at $p_L = 6 \ $GeV/c  (the experimental data \cite{pl6,plb}), and
  b) at $p_L = 11.75 \ $GeV/c
      (the experimental data \cite{plb,pl12})     }
\label{fig:1}       
\end{figure*}

The calculated analyzing power  at $p_L = 6 \ GeV/c$ is shown in
  Fig.1a Of course, in the original phenomenological analysis made in
  \cite{wak}
 all helicity amplitudes
 were used,
  but it can be seen that
  a good description, practically the same as in \cite{wak},
   of experimental data on the analyzing power
  can be reached only with
   one  hadron-spin flip amplitude.

\begin{figure*}
\resizebox{0.85\textwidth}{!}{%
 \phantom{.} \hspace{30mm} \includegraphics{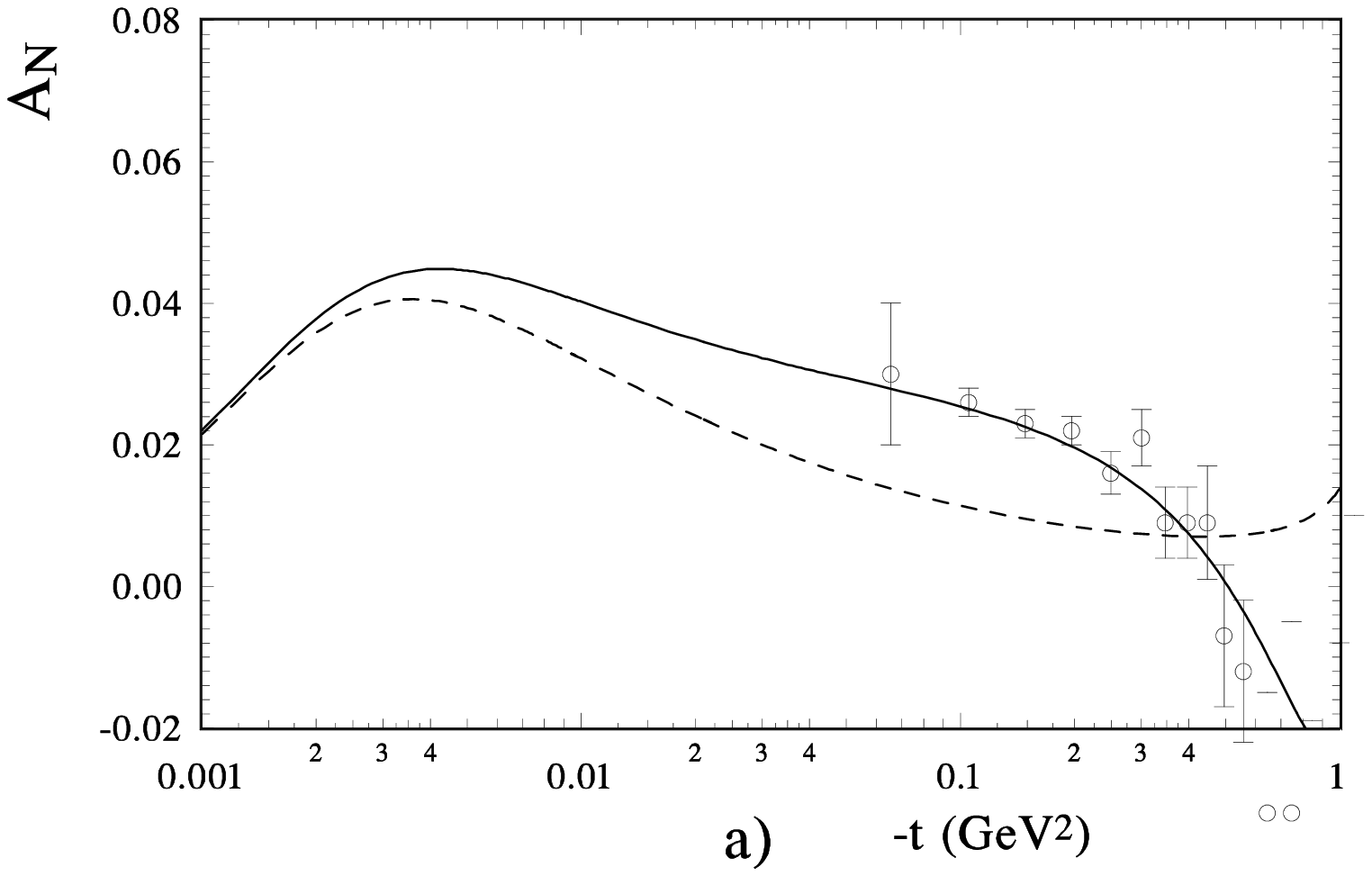}
  \includegraphics{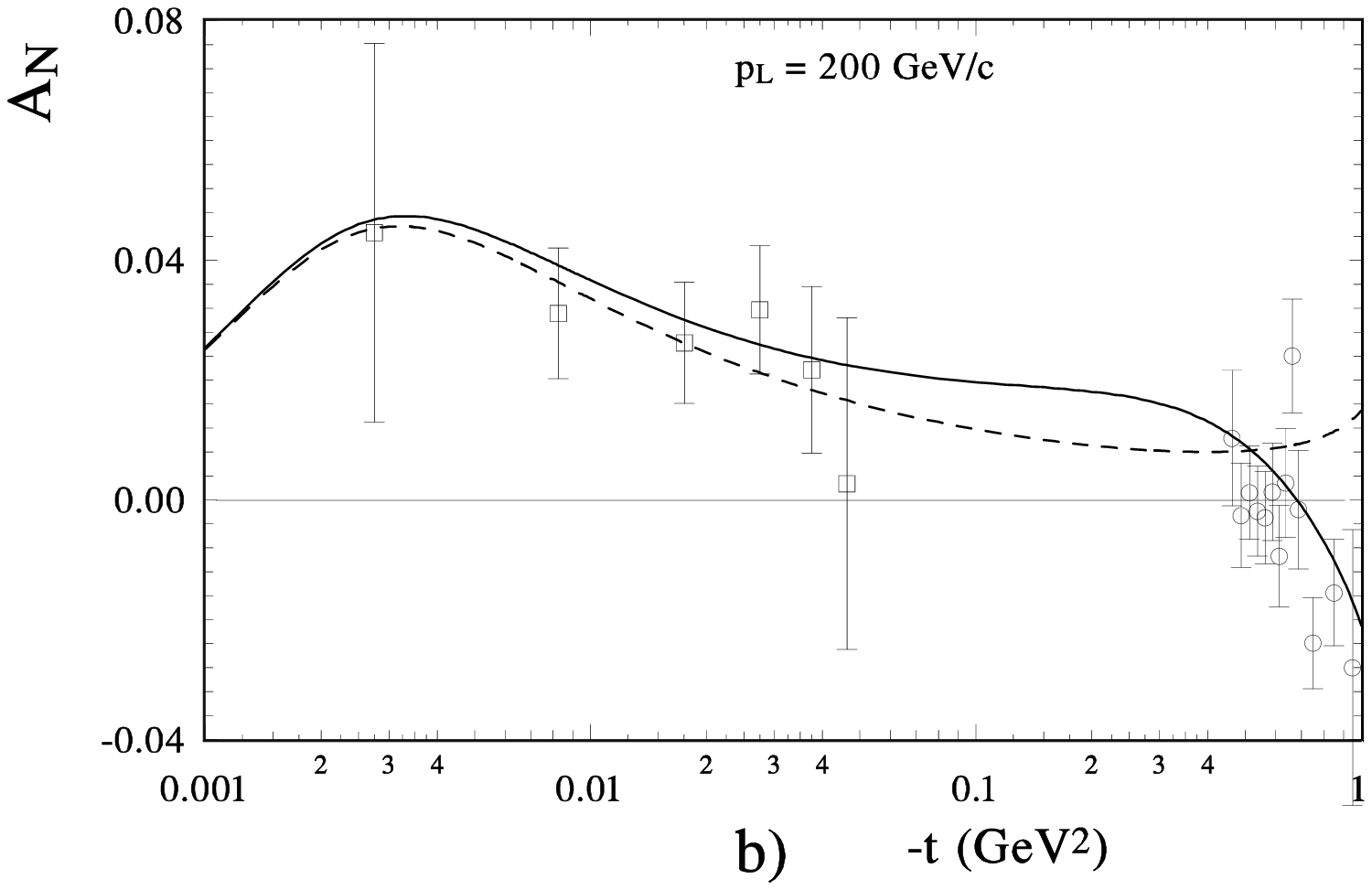}
}
\caption{The analyzing power $A_N$ of pp - scattering
      calculated:
   a) at $p_L = 45.5 \ $GeV/c (the experimental data \cite{p45}), and
   b)    at $p_L = 200 \ $GeV/c
       (the experimental data \cite{p200,b78})    }
\label{fig:1}       
\end{figure*}

    The experimental data at $p_L = 11.75 \ $GeV/c  seriously
 differ from those at $p_L = 6 \ $GeV/c but our calculations
 reproduce them sufficiently well (Fig.1b ).
 It is shown that our energy dependence
  was chosen correctly and we may hope that  further we will obtain
  correct values of the analyzing power.

       Really, our calculations at
  $p_L = 45.5 \ $GeV/c show a satisfactory description
  of the experimental data  (see Fig.2a).
   At this energy both of our
 parts of the amplitude give important contributions. The contributions
  to the analyzing power of the amplitudes (\ref{wak1}, \ref{wak2}) are
  approximately twice as large as the contributions of the model
  amplitudes.
  From Fig.2a we can see that in the region $|t| \approx
   0.2 \ GeV^2$ the contributions from the hadron spin-flip amplitudes
  are most important.

      At last, Fig.2b shows our calculations at $p_L = 200 \
  GeV/c$.  At this energy, the contributions of the phenomenological
  amplitudes are already very small and can be compared with the
  contributions of the model amplitudes only at $|t| = 0.5 \ GeV^2$
  where both the contributions are very small.

\begin{figure*}
\resizebox{0.85\textwidth}{!}{%
 \phantom{.} \hspace{30mm} \includegraphics{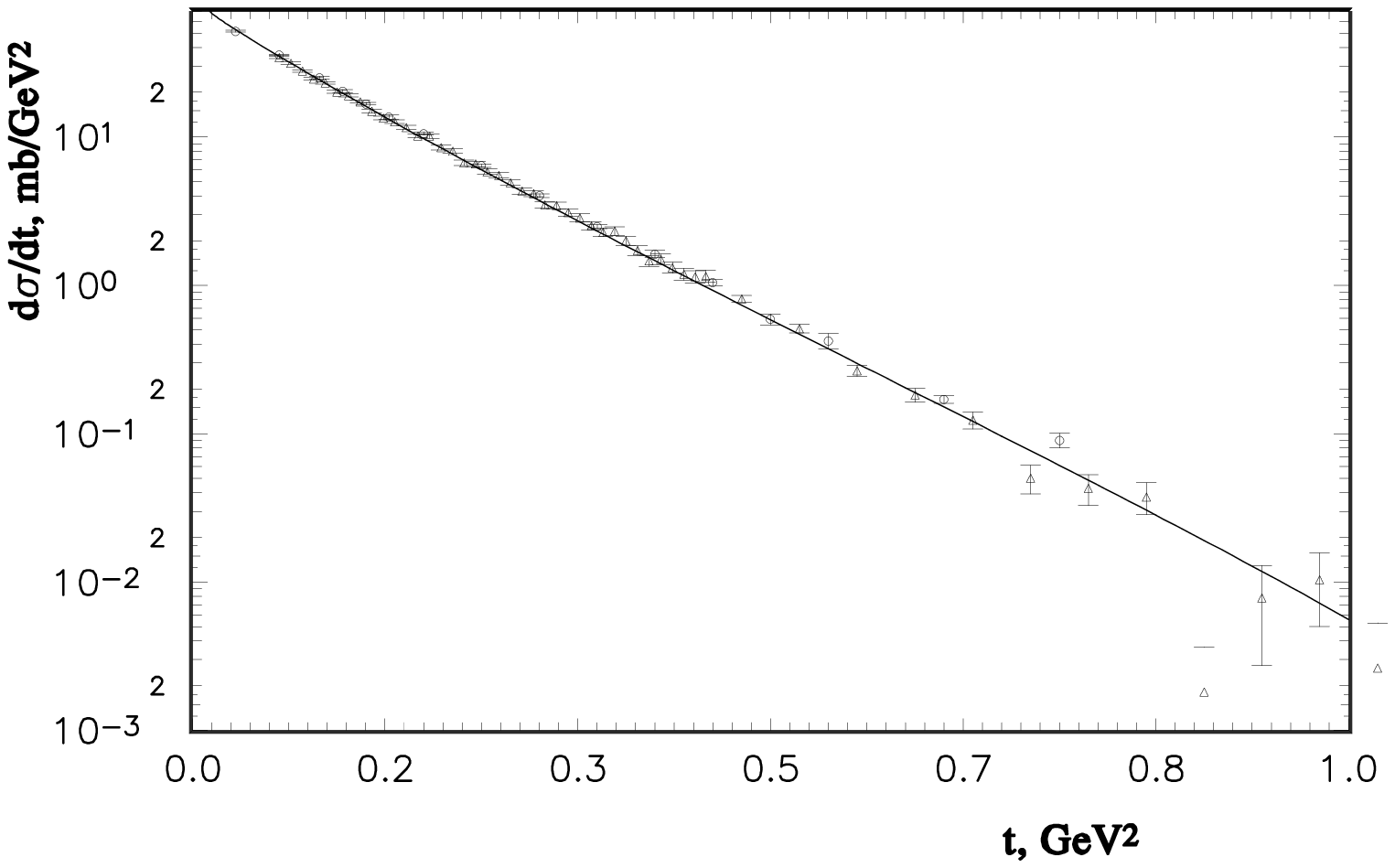}
  \includegraphics{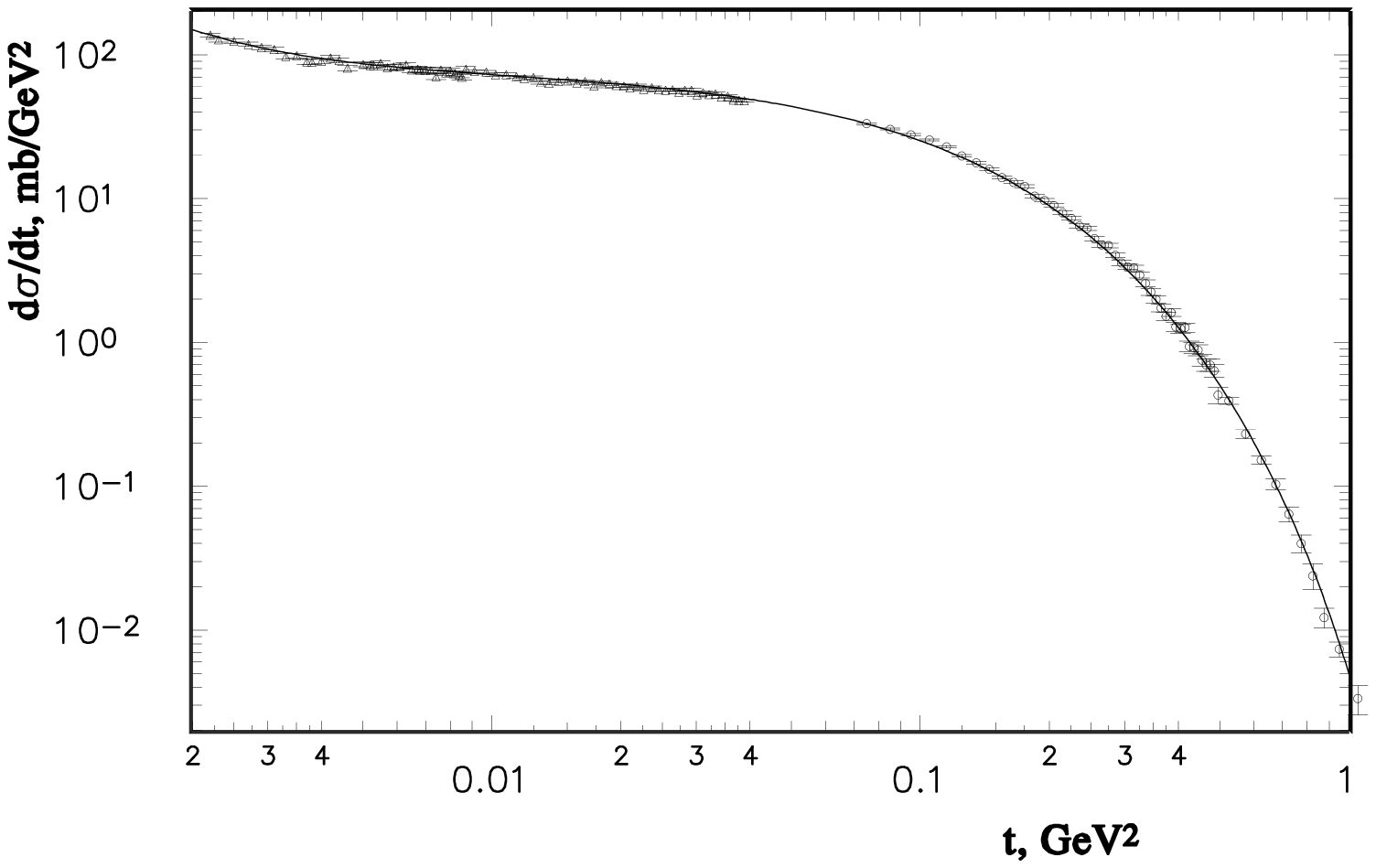}
}
\caption{The differential cross sections of pp - scattering
       calculated at $p_L = 50 \ $GeV/c [left] (the experimental data \cite{ds50ay,dsak}) and
   at $p_L = 100 \ $GeV/c [right]
       the experimental data \cite{dsak,ds100bu}}
\label{fig:1}       
\end{figure*}

   Let us check up how our amplitudes describe the differential cross sections, especially
   for intermediate  region where both the solutions give  one-order contributions.
   The calculations for $p_L = 50 \ $GeV/c and $p_L = 100 \ $GeV/c are presented on
   Fig. 3. The non-normalized experimental data \cite{dsak} were normalized to the
   experimental data \cite{ds50ay}  at $p_L = 50 \ $GeV/c and \cite{ds100bu} at $p_L = 100 \ $GeV/c.
   It is clear that the coincidence of
   the theoretical curves and experimental data  was obtained  sufficiently good for  both the
   energies and the whole examined region of the momentum transfer.
   We should like to emphasize  that
   we do not make a fit of the differential cross sections. We only sew the low and high energy
   solutions \cite{wak} and \cite{zpc}.
    The parameters of the factor-functions were chosen to obtain a qualitative description
    of the form of  $A_N$ at $p_L = 11.5 \ $GeV/c and then they were fixed.

   Note that we obtain a different energy dependence of
  the additional contributions $\Delta A_N$ to the pure $A_N^{CH}$ effect at
   different points of transfer momenta. The contribution at
  $|t|=0.1 \ GeV^2$ has a clear downfall with growing  $\sqrt{s}$,
  but in the range of maximum of $A_N^{CH}$ we have nearly constant
  contributions which are independent of energy.  So we cannot make
  the conclusion about energy dependence of $\Delta A_N$ at
  maximum of  $A_N^{CH}$
  measuring the energy dependence of the analyzing
  power at other points of the transfer momentum. However,
    it is one of
  the central points of many other analyses of the electromagnetic-hadron
  interference effect.

   The comparison of our calculations with the recent final experimental data
   obtained at RHIC \cite{an-hep} (see fig.4) at  $p_L = 100 \ $GeV/c shows
   suitable agreement. The preliminary experimental data were slightly above
   than the final ones and showed, on our opinion, the existence of the hadron spin flip
   contributions. The final data, say accurately, do not contradict such contributions.
   We will analyze  the final data in the next section.

       Especially note that
   it is very important to continue the measured range at the  largest
    momentum transfer.
   In future it is most important to measure $A_N$ in the range
   of the dip of the differential cross sections and high energies.
   The corresponding predictions were made in \cite{akch-rhic}.
   The value of $r_5 = R + iI$ - the ratio of the hadron spin-flip amplitude
 to the imaginary part of hadron
   spin-non-flip amplitude (without the kinematic factor $\sqrt{|t|}$)
  in \cite{akth} was obtained  in \cite{akth}   $R = -0.01 \pm 0.004 $ and $I= 0.082\pm 0.138$
   at energy region $p_L = 150-300$ GeV/c and
  $R = -0.037 \pm 0.022 $ and $I= 0.078\pm 0.182$
   at energy region $p_L = 45-205$ GeV/c.
 In our model calculations  at $p_L =100 \ $GeV and
  at the position of the maximum $A_N$ we obtain the value of
 $r_5(p_L=100 GeV, -t_{max}) = -0.015 - i 0.01 $.
    Of course, this value depends
   on both energy and transfer momentum. More complete
    analyses of these
   dependencies were carried out in \cite{cnizdr}.

\begin{figure}
\resizebox{0.45\textwidth}{!}{%
  \includegraphics{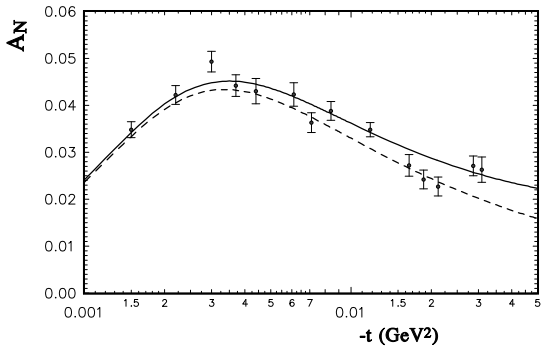}
}
\caption{The analyzing power $A_N$ of pp - scattering
      calculated at $p_L = 100 \ $GeV/c;
       (the full line is the model calculations;
       the dashed line is the model calculation of the $A_N^{CH}$;
}
\label{fig:1}       
\end{figure}

\section{Phenomenological analysis of $A_N^{CH}$}

   There is one important note. On Fig.4 our curve for
  pure electromagnetic-hadron interference  $A_N^{CH}$ reaches at  maximum
 the size  $4.37 \% $. On the contrary, in the  talks and publications,  the preliminary new
  experimental data are compared with the curve of $A_N^{CH}$ which
  reaches at its maximum approximately  $4.67 \% $.
  From the comparison of
  the curve with the new experimental data the authors made a conclusion
  that the contribution from the hadron-spin flip amplitude disappears.

  We study this problem to understand the contradiction with our calculations.
  Some authors suppose that the value $A_N^{CH}$ does not  practically depend on energy.
   In an early work \cite{k-l}, where the size of $A_N^{CH}$ was evaluated, it was obtained
  that
\ba
          A_N^{CH} \sim 4.5\% \ Im(a_h)/|a_h|,
\ea
  where $a_h$ is a spin-non-flip amplitude. In the case of a small real part of $a_h$
  this form leads to the   size of $A_N^{CH}$ independent of the size of $\sigma_{tot}$.
   However, this formula gives a small dependence of the size of $A_N^{CH}$
   on the $\rho(s,t)$ - the ratio of
  the real to imaginary part of the hadron spin-non-flip amplitude.
  Over a period of time  this short version
  of $A_N^{CH}$ was rewritten in the different forms which led to the different energy
   dependence. Our opinion is that when we calculate such a small correlations effect
   we have to take the complete form of $A_N^{CH}$, formula (\ref{an}). All the approximations
   must be reflected in the form of the helicity amplitudes and the size of the parameters.

  There is an important energy dependence
  which is connected with the
  energy dependence of the Coulomb-hadron interference term in the differential cross sections.
   This term is in most part proportional to  the size of  $\rho(s,t)$.
  The position of the maximum of
  the contribution of this term to the differential cross section at $t$
   coincides  approximately with the position of the maximum $A_N^{CH}$.


\begin{figure}
\resizebox{0.45\textwidth}{!}{%
  \includegraphics{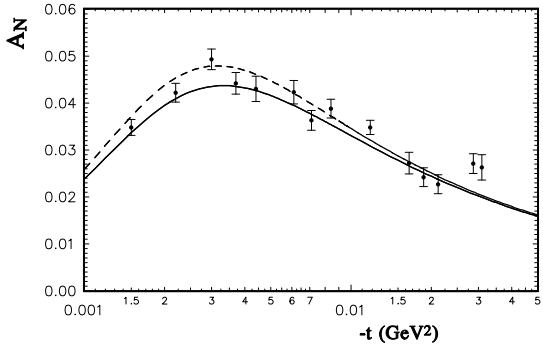}
}
\caption{  The phenomenological $A_N^{CH}$ of pp - scattering
       calculated at $p_L = 100 \ $GeV/c
       (the full and dashed lines are the calculations
        with $\rho=-0.1$ and with $\rho=0$ );
      the experimental data \cite{Bravar2}
}
\label{fig:6}       
\end{figure}

\begin{table*}
\caption{Table}
\label{tab:1}       
\begin{tabular}{lllllll}
\hline\noalign{\smallskip}
n & Form of $F_{i}(s,t)$  & $\sigma_{tot}(mb) $ & $\rho$ & $k_i$ & $k_r$ & $\sum_1^{14} \chi_i^2$ \\
\noalign{\smallskip}\hline\noalign{\smallskip}
1a & exponenential      & 38.46      & -0.105 & 0   & 0  & 30.62  \\
2a & exponential      & 38.46      & $-0.047\pm0.02$ & 0   & 0  & 23.22  \\
3a & exponential      & 36.5 $\pm$0.81     & -0.1 & 0   & 0  & 24  \\
4a & model      & 38.3      & -0.105 & 0   & 0  & 29.38  \\
\noalign{\smallskip}\hline\noalign{\smallskip}
1b & model      & 38.3      & -0.105 & -0.065   & -0.15  & 26.11  \\
2b & exponential      & 38.46      & -0.1 & $0.023\pm0.06$   & $-0.022\pm0.02$  & 20.64  \\
3b & exponential      & 38.46      & -0.1 & 0.02    & $0.023\pm0.01$  & 20.65  \\
4b & exponential      & 38.46      & -0.1 &$ 0.028\pm 0.01 $ & $0.$   & 20.95  \\
5b & exponential      & 38.46      & $-0.091\pm 0.032$ & 0.02    & $0.021\pm0.014$  & 20.58  \\
6b & exponential      & 38.46      & $-0.1\pm 0.03$ &$ 0.028 \pm 0.015$.   & $0.  $  & 20.58  \\
\noalign{\smallskip}\hline
\end{tabular}
\end{table*}

  Hence, the energy dependence of  $\rho(s,t)$ strongly impacts that
   of the maximum of  $A_N^{CH}$.
  In  our  semi-phenome-nological  descriptions we obtained the following
  values at $p_L =100 \ $GeV: \\
 $ \sigma_{tot} = 38.3 \ $mb;
 $ B(-t=0.003 \ GeV^{2} ) = 11.6 \ $GeV$^{-2}$;
 $B(-t=0.03 ) = 11.3 \ $GeV$^{-2}$;
 $\rho( -t=0.003  \ GeV^{2}) =-0.105$. \\
 The available  experimental data  (see \cite{shub}) are: \\
 $ \sigma_{tot} = 38.46 \pm 0.04 \ $mb;
 $B(-t=0.03 ) = 11.3 \ $GeV$^{-2}$;
 $\rho =-0.1 $.  \\
  So our values practically coincide with the existing experimental data.

  Of course, there also exists an energy dependence of the Coulomb-hadron phase
  which impacts  the size of the differential cross sections. In our original
  calculation we used this phase, which was obtained in \cite{selpht}
  with taking into account all correction factors. To check up
  the results, we used the simplest phase in the form of the West-Yennie \cite{w-y}
  and in the form of R. Cahn \cite{cahn}
\ba
  \varphi= -[ln(B |t|/2)+\gamma + ln(1 + 8/(B / \Lambda^2)). \label{ph-cn}
\ea
 where $B$ is the slope of the differential cross sections and $\Lambda^2=0.71$.
  The size of  $A_N^{CH}$  for our small momentum transfer region  changes only by $0.5 \% $.

 Now let us obtain the result for $A_N^{CH}$  with the simplest form
 of the hadron spin-non-flip amplitude:
\ba
  F_h(s,t) = \frac{\sigma_{tot}(s)}{4 \pi } \ (\rho +i)\  exp(B(s) \ t /2).
         \label{fhe1}
\ea
 and with the Coulomb-hadron phase of (\ref{ph-cn}).
 We take the hadron-spin flip amplitude in the form
\ba
   F_h^{sf}(s,t) =  \frac{\sigma_{tot}(s)}{4 \pi }\ (k_r \rho +i k_i) \ exp(B(s) \ t /2).
       \label{fhs}
\ea
    First, let us make the fit of the experimental data without the hadron-spin-flip
  amplitude. This case is presented in the upper part of Table  (1a-4a).
  The fit with parameters obtained in the model calculations but with the form
  of the amplitudes in the simple exponential form (\ref{fhe1}) is shown in the first row.
   If we take  $\rho$ as a free parameter,  $\chi^2$ essentially  decreases
    but the size of $\rho$ arrives at the value which strongly differs from the experimental
    data. The same situation is obtained if  $\sigma_{tot}$ is taken as a free parameter.
    The last row (5a) of the upper part of  Table  presents the calculation of $\chi^2$
      on the base of the model
    calculations but without the hadron spin-flip contribution. Note that we did not make
     the variation of the parameters of our model calculations for that case. The $\chi^2$
      was calculated by comparing the model calculations with the values of the experimental points.

    In the low part of Table  (1b-6b) the different fits with the existence of the hadron spin-flip
   amplitude are presented. Again the $chi^2$ on the base of the model
    calculations  with the hadron spin-flip contribution was calculated without the
    variations of the parameters. In this case,  $\chi^2$ decreased on the $4$ points.
   A more remarkable decrease in $\chi^2$ was obtained in the variations of the
   parameters of the hadron spin-flip amplitude for the cases of the exponential form of the helicity
   amplitudes.
   Of course, when both the parameters $k_r$ and $k_i$ are varied, the errors are large
    (see lines 2b in Table 1). If $k_r $ is fixed by some value or zero,
    the errors in the determination
   of the imaginary part of the hadron spin-flip amplitude are $30 \% $. It is to be note that
   the coefficient $k_r$ is multiplied by $\rho$ in the definition of the real part of the
   hadron spin-flip amplitude. Hence, the ratability between the imaginary and real parts
   $F_h^{nf}$ and $F_h^{sf}$ is practically the same but the signs are different, thus leading
   to the difference between the corresponding phases. As the fitting procedure shows,
   the small real part of $F_h^{sf}$ can be take with $k_r =0$. In this case, the $k_i$ grows
    (line 4b in  Table 1).  It is interesting that if we make the fit of  $\rho$
    and $k_i$ simultaneously, the size of  $\rho$ practically does not change (see line 6b
    and compare with line 2a in Table).

\section{Conclusion}

    The size of the parameters of the hadron spin-flip amplitude which can be obtained
   from the new experimental data at $p_L = 100 \ $ GeV/c
   are  determined with large errors.
   However,  $\chi^2$ decreases essentially. It is shown at least that the imaginary
   part of the hadron spin-flip amplitude differs from zero in this transfer momentum
   region and $p_L = 100 \ $ GeV/c. Note that the imaginary part of the spin-flip amplitude
   gives the contribution not only to the interference with the hadron spin-non-flip
    amplitude but also to the interference with the Coulombic part.
    Hence, any case, we cannot make a conclusion about
    the absence a contribution of the hadron spin flip amplitude at least on the base of
    these new experimental data.

   It is obvious from our analysis that  examining  the contributions
  of the hadron spin-flip amplitudes in the CNI effect using the experimental data
    in a wide energy region, one should
  take into account the energy dependence of all parts of the hadron
  scattering amplitude and its dependence on transfer momenta.
  Our descriptions of all available experimental data give
   about $3.5 \% $ of the predictions
  for RHIC energies    for the contributions of the hadron
  spin-flip amplitude to the maximum of the CNI effect. Of course,
  this estimation is very rough,
  but the comparison of the calculated $A_N$ and $A_N^{CH}$ with the new
   experimental data obtained at RHIC shows that at this energy the contribution of the
   hadron spin-flip amplitude is presented.
 More accurate estimations
  can be carried out only after a new experiment in this domain of transfer
  momenta at  higher energies and wider transfer momenta, especially
  in the dip region.

\vspace*{5mm}

 {\it Acknowledgements.} The author  would like to thank
  the Department of Theoretical and Mathematical Physics of the University of Liege for
 their hospitality and the FRNS for financial support.


\begin{thebibliography}{99}


\bibitem{swin} J. Schwinger Phys.Rev., {\bf 73} 407 (1948);
   L.I.Lapidus, Nucl. and Part., {\bf 9} 84 (1978);
   N.H.Battimore, E.Gotsman, E.Leader, Phis.Rev. {\bf D 18} 694 (1978).

\bibitem{sum} N.Akchurin, A.Bravar, M.Conte, A.Penzo,
   U of Iowa Report 93-04;
 W. Guryn {\it et al.}, {\it
Total and Differential Cross Sections and Polarization Effects
in pp Elastic Scattering at RHIC} (unpublished).
S.B.Nurushev, A.G.Ufimtsev, Proc. "Hera-N", Dubna, (1996);
 N.H. Battimore {\it et al.}, Phys.Rev D {\bf 59}  114010 (1999).

\bibitem{Bravar1} A. Bravar et al.,
  in Proceedings XVI International Symposium
   High Energy Spin Physics
   10-16 October (2004), Trieste,
 ed. Aulenbacher, Bradamante, Bressan, Martin, p.700;
  in Proceedings X International Conference ADS, Blois (2004);
  H. Okada et al. ibid, p.507,(nucl-ex/0502022).

\bibitem{Bravar2}
 A. Bravar et al., in Proceedings "X EDS International Conference",
   23-30 May, Blois, France (2005);
   A. Bravar et al., in Proceedings XIV International Conference
   High Energy Spin Physics
   27-30 September, Dubna (2004), ed. O. Teryaev, A. Efremov;
   W. Haeberli, Cern Courier, October 2005, p.15.

\bibitem{bsw}
  C. Bourrely, J. Soffer and T. T. Wu, Phys. Rev. D {\bf 19} 3249 (1979).


\bibitem{mog1}
 B.Z. Kopeliovich and B.G. Zakharov, Phys.lett. B 156  (1989);
 M. Anselmino and S. Forte, Phys. Rev. Lett. {\bf 71}, 223 (1993) ;
   S.V. Goloskokov, Phys. Lett. B {\bf 315}  459 (1993);
  A. E. Dorokhov, N. I. Kochelev and Yu. A. Zubov,
Int. Jour. Mod. Phys. {\bf A8},  603 (1993);


\bi{zpc} S.V.Goloskokov, S.P.Kuleshov, O.V.Selyugin,
    Z.Phys. C {\bf 50} 455 (1991).


\bibitem{selpht} O.V. Selyugin,
Int. Jour. Mod. Phys. {\bf A 12} 1379 (1997);

\bibitem{akex} N.Akchurin, et al., Phys.Rev. {\bf D 48}, 326
   (1993).
\bibitem{akth} N.Akchurin, N.H.Buttimore, A.Penzo, Phys.Rev. {\bf D 51},
     3944 (1995).





\bibitem{trosh} A.D.Krisch, S.M.Troshin, hep-ph/9610537.


\bibitem{soff} C.Bourrely, J.Soffer, hep-ph/9611234.


\bibitem{btt96} N.H. Buttimore, Proc. XII High Energy Spin Physics
   10-14 September, Amsterdam (1996).

\bibitem{berg}  E.L. Berger, A.C. Irving, C. Sorensen, Phys.Rev. D {\bf 17} (1978) 2971.

\bi{wak} M.Sawamoto, S.Wakaizumi, Proc Theor.Phis. {\bf 62} 1293 (1979).


\bibitem{sum-L} N.H. Buttimore et al.,   Phys.Rev. D {\bf 59} (1999) 114010.




\bibitem{M-Pred} A.F. Martin, and E. Predazzi, Phys. Rev. D  {\bf 66} (2002) 034029;
                 E. Predazzi and O.V. Selyugin, Eur.Phys.J. A {\bf 13} (2002) 471.

\bibitem{cnizdr} O.V. Selyugin, {\it Physics of Atomic Nuclei} {\bf 62 } 333 (1999).

\bibitem{gaz} S. Gasiorowicz, "Elementary particle physics", John Wiley \& Soms,Inc,
   New York - London - Sydney.

\bibitem{pred} V. Barone, E. Predazzi, in book "High Energy particle Diffraction", NY (2002).

\bibitem{tru-an-t} T.L. Trueman, hep-ph/9610429.

\bi{g2}  S.V. Goloskokov, S.P. Kuleshov, O.V. Selyugin,
      Yad.Fiz., {\bf 46}, 597 (1987).

\bi{g4}  S.V. Goloskokov, S.P. Kuleshov, O.V. Selyugin,
       Yad.Fiz., {\bf 52}, 561 (1990).
-94", Uzhgorod, p.97 (1994).






\bi{pl6} D.  Miller {\it et al.} Phys.Rev. 1977. V.D16. P.2016;
    M.  Borghini  {\it et al.},   Phys.Lett. B {\bf 31} ( 1970) 405 ;
     R. Diebold  {\it et al.},    Phys.Rev.Lett.  {\bf 35} (1975) 632;
     D.R. Rust  {\it et al.},      Phys.Lett. B {\bf 58} (1975) 114;
     R.D. Klem {\it et al.},      Phys.Rev. D {\bf 15} (1977)  602;
    J.R. O'Fallon  {\it  et al.},   Phys.Rev.  D {\bf 17} (1978) 24.
\bi{plb} M. Borghini  {\it et al.},  Phys.Lett. B {\bf 24} ( 1966) 77.
\bi{pl12} S.L.  Kramer {\it et al.},   Phys.Rev. D {\bf 17} ( 1978)1709;
     K. Abe  {\it et al.},     Phys.Lett. B {\bf 63} (1976) 239.

\bi{p45} A. Gaudot {\it et al.}   Phys.Lett. 1976. V.B61. P.103.
\bi{p200}  G. Fidegaro {\it et al.}   Phys.Rev. 1981. V.B105. P.309.
\bi{b78}  E.L. Berger {\it et al.},   Phys.Rev. D {\bf 17} (1978) 2971.


\bibitem{ds50ay} D.S. Ayres, {\it et al.}, Phys.Rev. D {\bf 15}(1977) 3105.

\bibitem{dsak} C. W. Akerlof,  {\it et al.}, Phys.Rev. D {\bf 14}(1976) 2864.


\bibitem{ds100bu} J.P. Burq, {\it et al.}, Nucl.Phys. B {\bf 217}(1983) 285.


\bibitem{an-hep} H. Okada, et al., hep-ph/0601001.

\bibitem{akch-rhic} N.Akchurin, S.V. Goloskokov, O.V. selyugin, Int. J. Mod. Phys. A {\bf 14}
   (1999) 252.


\bibitem{k-l}
 A.P. Vanzha, L.I. Lapidus, and A.V. Tarasov,  Yad.Fiz. {\bf 16} (1972)1023;
 B.Z. Kopeliovich and L.I. Lapidus, Yad.Fiz. {\bf 19} (1974)218.
              [Sov.J.Nucl.Phys. {\bf 19} (1974) 114.


\bibitem{w-y} G.B. West and D.R. Yennie, Phys.Rev., {\bf 172} (1968) 1414.

\bibitem{cahn} R.N. Cahn, Z.Phys. C {\bf 15} (1982) 253.




\bibitem{shub} Landolt-Bornstein, New Series, {\bf 9} New York (1980).






\end{thebibliography}
\end{document}